\documentclass{aa}  
\usepackage{graphicx}
\usepackage{txfonts}
\usepackage{aalongtable}
\begin{document}
  \title{Complete identification of the Parkes half-Jansky sample of GHz peaked spectrum radio galaxies\thanks{Based on observations collected at the European Southern Observatory Very Large Telescope, Paranal, Chile (ESO prog. ID No. 073.B-0289(B)) and the European Southern Observatory 3.6m Telescope, La Silla, Chile (prog. ID No. 073.B-0289(A)).}}

  \author{N. de Vries \inst{1}
    \and
    I. A. G. Snellen \inst{1}
    \and
    R. T. Schilizzi \inst{1,2}
    \and
    M. D. Lehnert \inst{3,4}
    \and
    M. N. Bremer \inst{5}
  }

  \offprints{vriesn@strw.leidenuniv.nl}

  \institute{Leiden Observatory, Leiden University, P.O. Box 9513, 2300 RA, Leiden, The Netherlands\\
                  \email{vriesn@strw.leidenuniv.nl}
    \and
    International Square Kilometre Array Project office, Postbus 2, 7990 AA, Dwingeloo, The Netherlands
    \and
    Max-Planck-Institut f\"ur extraterrestrische Physik (MPE), Postfach 1312, 85741 Garching, Germany
    \and
    Laboratoire d'Etudes des Galaxies, Etoiles, Physique et Instrumentation, Observatoire de Paris, 5 Place Jules Janssen, 92195 Meudon, France
    \and
    Department of Physics, Bristol University, H H Wills Physics Laboratory, Tyndall Avenue, Bristol BS8 1TL
  }

   \date{Received 4 October 2006 /
          Accepted 15 December 2006}

  \abstract
    {Gigahertz Peaked Spectrum (GPS) radio galaxies are generally thought to be the young counterparts of classical extended radio sources. Statistically complete samples of GPS sources are vital for studying the early evolution of radio-loud AGN and the trigger of their nuclear activity. The `Parkes half-Jansky' sample of GPS radio galaxies is such a sample, representing the southern counterpart of the 1998 Stanghellini sample of bright GPS sources.}
   {As a first step of the investigation of the sample, the host galaxies need to be identified and their redshifts determined.}
   {Deep $R$-band VLT-FORS1 and ESO 3.6m EFOSC II images and long slit spectra  have been taken for the unidentified sources in the sample.}
   {We have identified all twelve previously unknown host galaxies of the radio sources in the sample. Eleven have host galaxies in the range 21.0 $< R_C <$ 23.0, while one object, PKS~J0210+0419, is identified in the near infrared with a galaxy with $K_s$ = 18.3. The redshifts of 21 host galaxies have been determined in the range 0.474 $< z <$ 1.539, bringing the total number of redshifts to 39 (80\%). Analysis of the absolute magnitudes of the GPS host galaxies show that at $z>1$ they are on average a magnitude fainter than classical 3C radio galaxies, as found in earlier studies. However their restframe UV luminosities indicate that there is an extra light contribution from the AGN, or from a population of young stars.}
   {}


   \maketitle
%
%
\section{Introduction}
Gigahertz Peaked Spectrum (GPS) radio sources are among the brightest radio sources in the sky. They are compact objects characterized by a turnover in their radio spectra at about 1 GHz in frequency. Their radio morphologies are small-scale versions of the well known extended Fanaroff \& Riley (FR) I/II radio sources, but with a physical extent of only 10-100 pc, well within the central regions of their host galaxies (Stanghellini et al. \cite{Stanghellini99}). To explain the compactness of these sources two scenarios were proposed: (1) these objects are very young radio-loud active galaxies which may evolve into kpc-sized Compact Steep Spectrum (CSS) sources and eventually grow to become FR I/IIs, (2) they are `frustrated' radio sources, millions of years old, but confined by a dense interstellar medium (Baum et al. \cite{Baum90}). In recent years compelling evidence has accumulated in favour of the youth scenario. VLBI monitoring of the archetype GPS sources, that began in the early 1980s,  has now conclusively shown these sources expand in size, implying source ages of 10$^{2-3}$ years only (Owsianik et al. \cite{Owsianikb}; Owsianik \& Conway \cite{Owsianika}; Tschager et al. \cite{Tschager}; Polatidis \& Conway \cite{Polatidis}), in good agreement with spectral age estimates (Murgia \cite{Murgia}). Recently Vink et al. (\cite{Vink}) have shown that their optical line emission is relatively underluminous, exhibiting a possible trend with radio source age. This is consistent with the fact that the Str\"omgren sphere should still be expanding in these objects, and that we are witnessing the birth of their narrow emission line regions. If indeed GPS galaxies are young, as now seems to be very likely, they form the key objects to study the early evolution of powerful radio-loud AGN, and the trigger of nuclear activity. In this paper we present new optical observations of a sample of GPS galaxies, aimed at identifying all host galaxies and determining their redshifts. Section~\ref{Ssam} defines this sample, section~\ref{Sobs} describes the observations, and the results are presented and discussed in Sect.~\ref{Sres}.
%
%
\section{The Parkes half-Jansky sample of GPS galaxies} \label{Ssam}
Snellen et al. (\cite{Snellen02}) have defined a sample of bright GPS sources in the southern/equatorial sky, representing the counterpart of the sample of Stanghellini et al. (\cite{Stanghellini98}), although somewhat deeper. High flux density GPS sources are rare and therefore all sky coverage is needed in order to obtain a statistically significant number of sources. Furthermore, access to large optical telescopes is better for the southern hemisphere, at least for European astronomers. The selection criteria of the sample are described in detail in a previous paper (Snellen et al. \cite{Snellen02}). Summarizing, it consists of 49 sources selected from the Parkes multifrequency survey (PKSCAT90, Wright \& Otrupcek \cite{Wright}), with $-40\degr < \delta < + 15\degr$, $|b| > 20\degr$ and $\rm{S_{2.7 GHz} > 0.5 Jy}$. The sample only consists of GPS radio sources associated with galaxies since GPS quasars do not seem to be related to GPS galaxies, despite having similar radio characteristics, and may not be young (e.g. Snellen et al. \cite{Snellen99}). Before the current work, 75\% of the radio sources had been optically identified with a host galaxy and 40\% had known redshifts.
%
%
\section{Observations} \label{Sobs}
\subsection{VLT-FORS1 spectroscopy}
Optical long slit spectroscopy was performed on objects in the sample using the European Southern Observatory (ESO) Very Large Telescope (VLT) at Paranal, in Chile, from March to September 2004. All observations were obtained in service mode with the visual and near UV FOcal Reducer and low dispersion Spectrograph (FORS1), using exposure times of 1800 s with the grism GRIS-300V (ESO \# 10) in combination with the order separation filter GG 435. The FORS1 has a Tektronix 2048~$\times$~2048 CCD detector with 24$\mu$m pixels, resulting in a scale of 0.2\arcsec/pixel (with the Standard Resolution collimator) and a slit length of 6.8\arcmin. The grism results in a dispersion of 2.69 \AA/pixel and a total wavelength coverage of 4450 - 8650 \AA. The spectra were taken using a slit width of 1.0\arcsec, resulting in a spectral resolution of FWHM = 5 pixels (13 \AA). Usually the slit was oriented near the parallactic angle, unless a second interesting object was located near the GPS galaxy (e.g. a possibly associated companion galaxy). The data reduction was carried out in a standard way using NOAO's IRAF reduction software. One dimensional spectra were extracted by summing in the spatial direction over an aperture as large as the spatial extent of the continuum or the brightest emission line. Acquisition images were taken to center the slit and in addition, for a few sources, to determine their $R$-band magnitude.
\subsection{ESO 3.6m observations}
Optical CCD imaging and spectroscopy were also performed using the ESO 3.6m Telescope at La Silla, in Chile, on March 23 and 24, 2004. For all observations we used the ESO Faint Object Spectrograph and Camera (EFOSC II), which has a 2048~$\times$~2048 CCD detector with 15$\mu$m pixels resulting in a scale of 0.157\arcsec/pixel and a field size of 5.4\arcmin~$\times$~5.4\arcmin. Spectroscopic observations were carried out in long slit mode with a slit width of 1.2\arcsec, using the EFOSC grism \#6, which has a wavelength coverage of 3860 - 8070 \AA\ and a dispersion of 137 \AA/mm or 2.06 \AA/pixel. The slit was always oriented near the paralactic angle. The reduction of the spectra was carried out in a similar way to the VLT data.

Photometric observations were carried out in Gunn $r$-band (EFOSC filter r\#786). The reduction of the images was carried out in a standard way using NOAO's IRAF reduction software. Astrometry was performed using data from the USNO-B1.0 Catalog, extracted with the VizieR catalogue access tool. For each image, catalogued positions of at least ten stars were used. With the IRAF procedure CCMAP, the equatorial coordinates were determined with errors always well within one pixel. Optical identifications were found, within the 1-$\sigma$ uncertainty ellipse, for all seven radio sources. Photometry was carried out using the IDL procedure APER with the aperture diameter set to 2.51\arcsec\ (16 pixels). The magnitude scale was calibrated using Landolt standard stars, with their Cousins $R_C$ magnitudes converted to Gunn $r$ using the conversion formula from Schombert et al. (\cite{Schombert}):
\begin{equation}
  r = R_C + 0.280 +0.038 \left( R_C - I_C \right)
\end{equation}
For better comparison with the rest of the sample, we also calculated Cousins $R_C$ magnitudes of the sources by using the unconverted Cousins $R_C$ magnitudes of the standard stars to calibrate our Gunn $r$ data. Since the colors of the sources are unknown, this conversion introduces extra uncertainties. However, these were always found to be small compared to the photometric errors.
\subsection{Additional near-infrared photometry}
Earlier optical observations of the GPS source PKS~J0210+0419 (Snellen et al. \cite{Snellen02}), did not result in an identification, with a lower limit of $m_R > 24.1$. We therefore observed this source in $K_s$-band using the SOFI near-infrared camera on ESO's New Technology Telescope (NTT). SOFI is equipped with a Rockwell $1024 \times 1024$ detector that provides images with a pixel scale of $0.288\arcsec$/pixel, and a field of view of about $4.9\arcmin \times 4.9\arcmin$ (Large Field imaging mode). The SOFI $K_s$ filter has a central wavelength of 2.162 $\mu m$ and a width of 0.275 $\mu m$. Details of the $K_s$-band observations of PKS~J0210+0419 and of other sources in the sample will be published in a following paper, however we present the $K_s$-band identification and magnitude of PKS~J0210+0419 ($K_s$ = 18.3) here for completeness.
%
%
\section{Results and discussion} \label{Sres}
\begin{figure}[ht]
\centering
 \includegraphics[width=8.5cm]{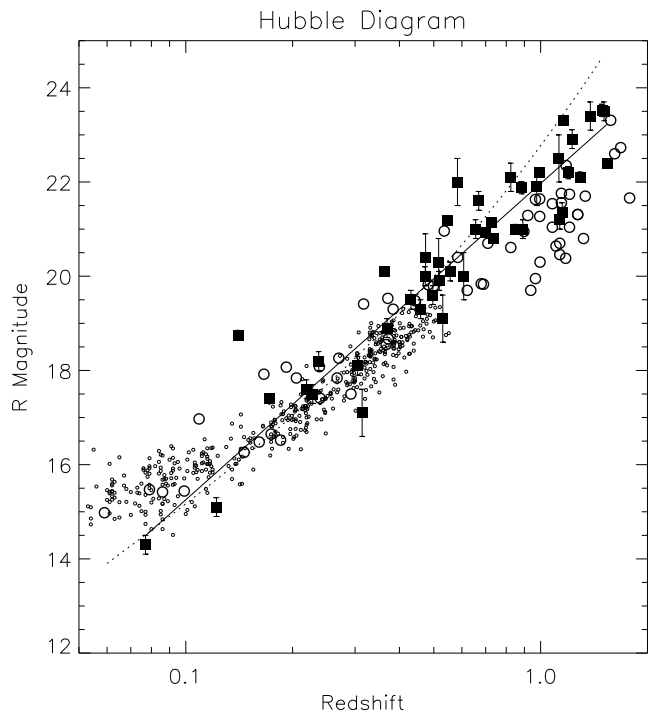}
\caption{Cousins $R_C$ Hubble diagram of GPS galaxies (solid squares), 3C galaxies (open circles), and LRGs (small circles). The dotted line is the $R$-band Hubble relation as derived by Snellen et al. (\cite{Snellen96a}), the solid line is a linear least squares fit to the current data.}
\label{Hubble}
\end{figure}
In total, eleven previously unidentified sources in the sample have been observed in $R$-band with the ESO 3.6m and VLT telescopes. All were optically identified, with 21.0 $< R_C <$ 23.0. Including the one identification in $K$-band (PKS~J0210+0419), this means that the Parkes half-Jansky sample of GPS galaxies is now {\sl completely} identified. Note that we have left one source (PKS~J1600$-$0037) out of our statistical sample, because it is located too close to a 12$^{\mathrm{th}}$ magnitude star to make an optical identification possible. Since this is a random occurence, it does not introduce any selection effects, so the sample will remain statistically complete. The results of the photometric observations are given in Table~\ref{photdet}; in column 1 the source name, in column 2 the exposure time, in column 3 the observed Gunn $r$-band magnitude with its error, and in column 4 the observed or deduced Cousins $R_C$-band magnitude.

We have also taken deep spectra of 24 objects in total. These have resulted in twenty secure redshifts, based on two or more emission or absorption lines, all in the range 0.474 $< z <$ 1.539. One spectrum (PKS~J1556-0622) resulted in a tentative redshift ($z=1.195$), based on only one line. We assumed the line to be [OII] 3727 \AA, because of the resemblance between this spectrum and those of PKS~J2339$-$0604 and PKS~J2212+0152, and the absence of plausible alternatives. A further nine sources in the sample (19\%) remain without redshift. The results of the spectroscopic observations are given in the appendix (Table~\ref{specdet}); in column 1 the source name, in column 2 the exposure time, in column 3 the name of the telescope used, in columns 4-7 respectively the name, rest wavelength, observed wavelength and rest-frame equivalent width of the identified spectral features, and in column 8 the derived redshift for each identified spectral feature and for each source. All spectra are also shown in the appendix (Fig.~\ref{Figspec}).
\begin{table}
\caption{Details of the photometric observations.}
\label{photdet}
\centering
\begin{tabular}{c r c c}
\hline
\multicolumn{4}{c}{ESO 3.6m Telescope observations} \\
\hline \hline
Object name	& t (s)	& $m_r$		& $m_{R_C}$ \\
\hline
PKS~J0441$-$3340	&  600	& 22.8 $\pm$ 0.4	& 22.5 $\pm$ 0.4 \\
PKS~J0913+1454	& 1200	& 23.2 $\pm$ 0.5	& 22.9 $\pm$ 0.5 \\
PKS~J1044$-$2712	& 1200	& 23.1 $\pm$ 0.4	& 22.8 $\pm$ 0.4 \\
PKS~J1057+0012	& 1200	& 22.6 $\pm$ 0.3	& 22.3 $\pm$ 0.3 \\
PKS~J1109+1043	& 1200	& 22.9 $\pm$ 0.3	& 22.6 $\pm$ 0.3 \\
PKS~J1122$-$2742	& 1600	& 23.3 $\pm$ 0.5	& 23.0 $\pm$ 0.5 \\
PKS~J1135$-$0021	&   60	&		& 21.9 $\pm$ 0.4 \\
PKS~J1648+0242	& 1200	& 22.4 $\pm$ 0.3	& 22.1 $\pm$ 0.3 \\
\hline
\multicolumn{4}{c}{ESO VLT observations} \\
\hline \hline
Object name	& t (s)	&	&  $m_{R_C}$ \\
\hline
PKS~J0401$-$2921	& 10	& 	&  21.0 $\pm$ 0.2 \\
PKS~J1345$-$3015	& 10	& 	&  21.2 $\pm$ 0.2 \\
PKS~J1352+1107	& 10	& 	&  21.0 $\pm$ 0.2 \\
\hline
\multicolumn{4}{c}{ESO NTT observation} \\
\hline \hline
Object name	& t (s)	& \multicolumn{2}{c}{$m_{Ks}$} \\
\hline
PKS~J0210+0419	& 450	&  \multicolumn{2}{c}{18.3 $\pm$ 0.2} \\
\hline
\end{tabular}
\end{table}
\begin{figure*}[th] 
\centering
\hbox{
 \includegraphics[width=8.5cm]{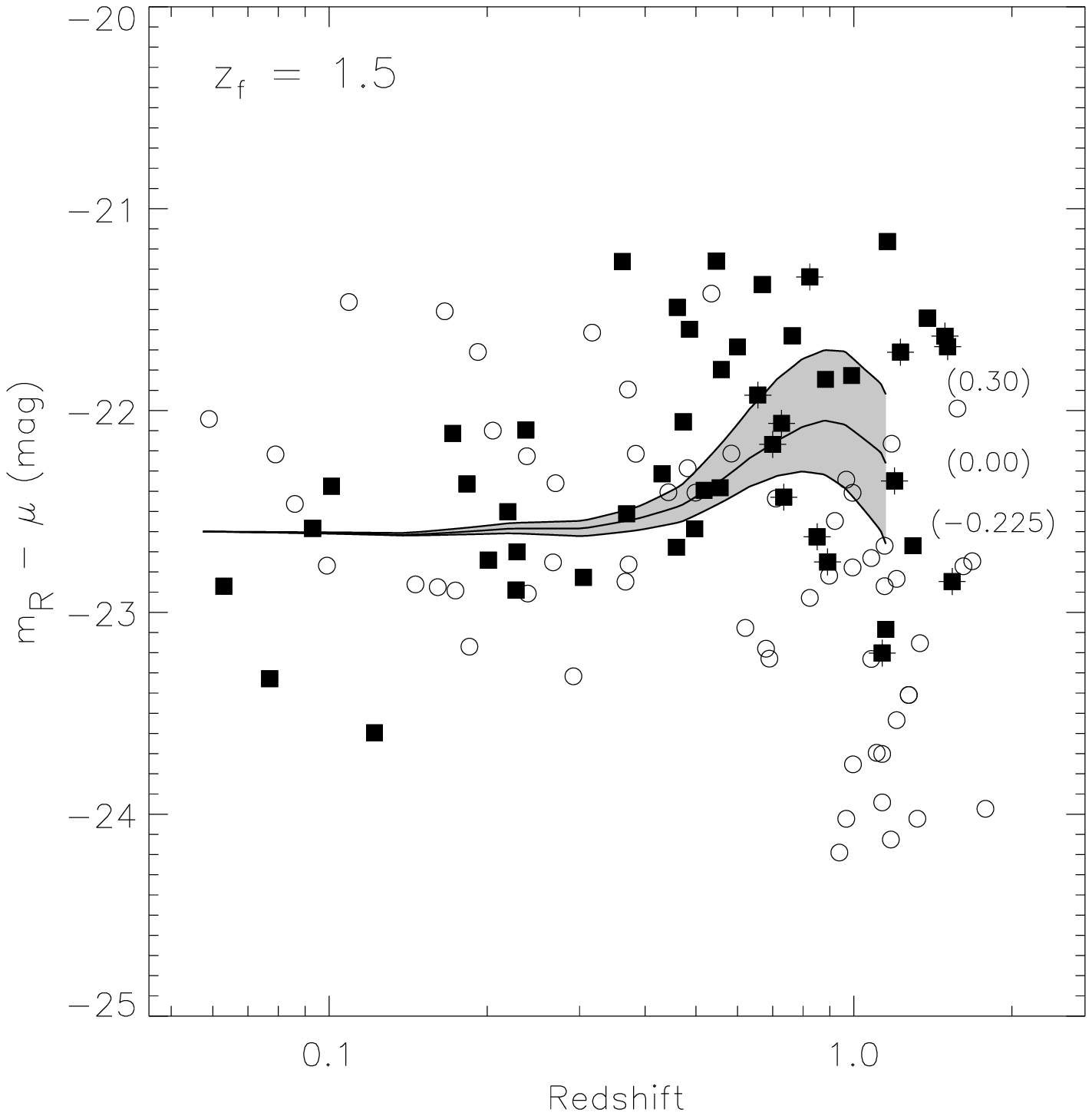}
 \includegraphics[width=8.5cm]{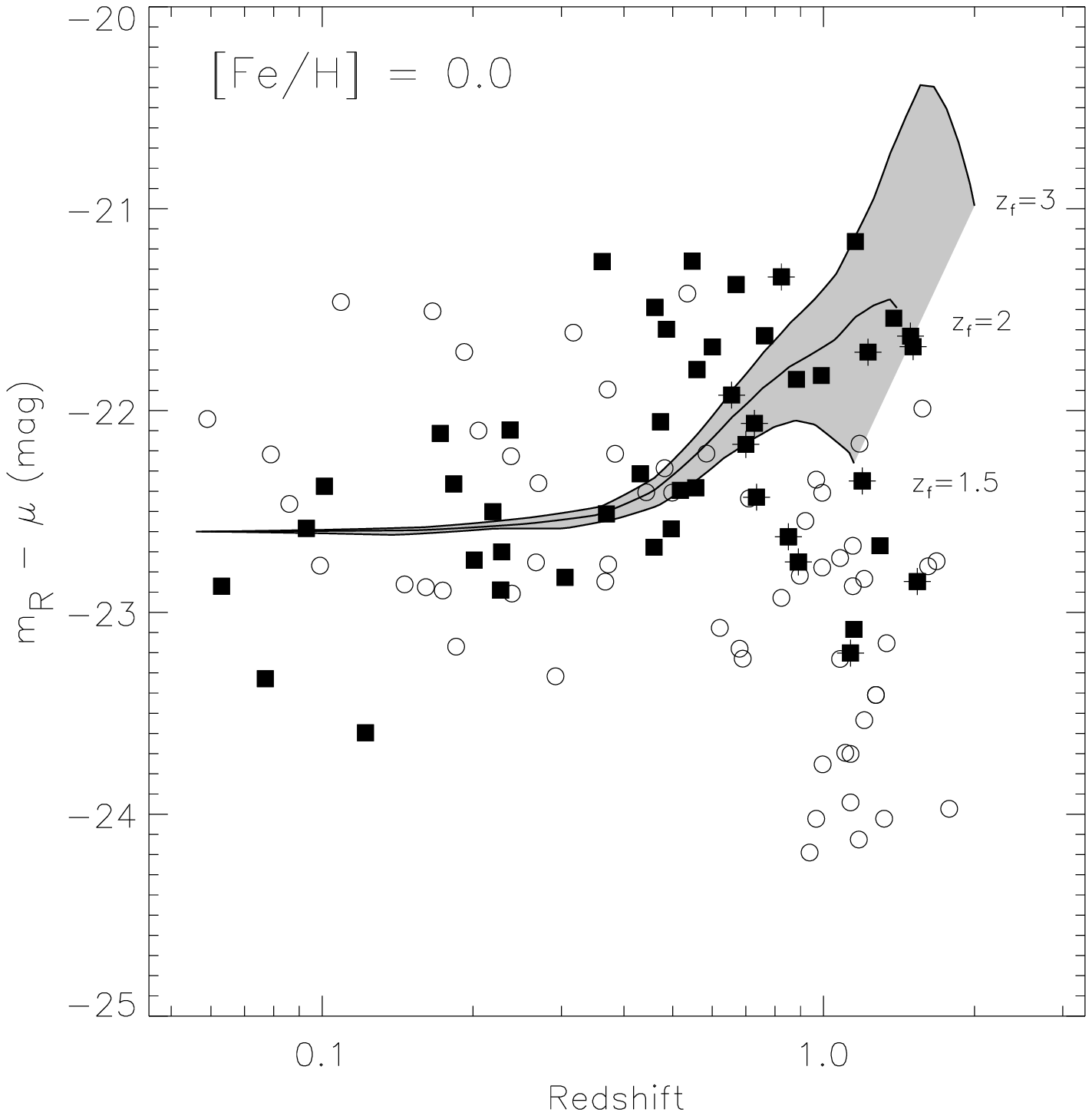}
}
\caption{(Left) Absolute magnitudes (without k-correction, $m_R-\mu$) for the combined samples of GPS galaxies (solid squares) and 3C galaxies (open circles). The grey shaded area is the modelled evolution, assuming a formation redshift of $1.5$ and a range of metallicities from $-0.225$ to $0.30$. Evolutionary tracks for the extreme metallicities as well as one for solar metallicity are overplotted, with metallicity decreasing from the upper to the lower track. (Right) The same figure, except that the evolution models now vary in formation redshift and keep the metallicity fixed at solar. The evolutionary tracks are for a formation redshift of 1.5 (lower), 2 (middle) and 3 (upper).}
\label{FigEvol}
\end{figure*}
\subsection{R-band Hubble diagram} \label{SSHub}
Now that we have significantly increased the number of GPS galaxies with known redshifts, particularly around redshift $z \sim 1$, it is interesting to review the R-band Hubble relation as previously discussed by Snellen et al. (\cite{Snellen96a}) and O'Dea et al. (\cite{O'Dea96}). For this purpose we combined our southern/equatorial sample with the northern sample of GPS sources of Stanghellini et al. (\cite{Stanghellini98}), excluding all those objects identified with quasars. Spectroscopic observations of two sources from our sample (PKS~J1203+0414 and PKS~J1506$-$0919) reveal broad emission lines ($\sim$ 10$^4$ km/s) and non-thermal emission. Technically these sources are now identified as quasars, and have been omitted from the sample. The resulting $R$-band Hubble diagram of GPS galaxies (solid squares) is shown in Fig.~\ref{Hubble}. For comparison, 3C radio galaxies (open circles)  from a compilation of samples (Best et al. (\cite{Best}), Eales (\cite{Eales}), and de Koff et al. (\cite{Koff})) and a subset of the `Luminous Red Galaxies' (LRGs; small circles) sample (Eisenstein et al. \cite{Eisenstein}) from the Sloan Digital Sky Survey (SDSS) are shown. If necessary, the magnitudes were converted to `total magnitudes' in Cousins $R_C$. Note the one GPS source (PKS~J1604$-$2223, with $R_C = 18.75$ and $z = 0.141$) that is over two magnitudes fainter than the general population. This is much too faint for a typical powerful AGN host galaxy, for which a number of explanations could be given. Although the source is located in a region with high galactic extinction, it is unlikely that this is the reason for the offset, since the object is also over two magnitudes fainter in $K$-band (de Vries et al. in prep.) than expected from the $K$-band Hubble diagram (as presented by Snellen et al. \cite{Snellen96b}). Of course there is always the possibility that a foreground galaxy is located between us and the radio source. This would also explain why the optical spectrum shows no emission lines or signs of nuclear activity. Alternatively, it could be that the source is not a typical young powerful AGN, but some other object that happens to have a similar radio spectrum. Although PKS~J1604$-$2223 is shown in Fig.~\ref{Hubble}, it is left out of any further analysis.

Now that the $R$-band Hubble relation is better sampled at $z \sim 1$ it is clear that the new data points systematically fall below the original fit (dotted line) of Snellen et al. (\cite{Snellen96a}). Since this relation is often used to estimate redshifts of GPS galaxies for which only photometric $R$-band data are available, it is valuable to determine a relation that holds for a larger redshift range, out to $z \simeq 1.5$.
We performed a linear least squares fit to the current data and found the relation:
\begin{equation}
  m_{R_C} = 21.97 + 6.71 \times \log(z)
\end{equation}
with a dispersion of 0.6 magnitudes. To estimate redshifts of GPS galaxies, this relation can be inverted to give:
\begin{equation}
  z_{est} = 5.32 \cdot 10^{-4} \times (1.41)^{m_R}, \quad \sigma_z \simeq 0.21 \times z_{est}
\end{equation}
with $\sigma_z$ the 1-$\sigma$ uncertainty in the redshift estimate.

Furthermore the new data at $z \sim 1$ confirm that GPS galaxies are on average 1.0 magnitudes fainter in this redshift range than 3C radio galaxies, as was initially claimed by Snellen et al. (\cite{Snellen96a}). This agrees with the hypothesis that GPS galaxies are redder, due to the lack of the extra, blue light associated with the radio-optical alignment effect (Chambers, Miley \& van Breugel \cite{Chambers}; McCarthy et al. \cite{McCarthy}; Best, Longair \& R\"ottgering \cite{Best}). The LRGs form a volume limited sample of the most luminous, intrinsically red galaxies out to $z \simeq 0.55$. They are selected on the basis of colors and magnitudes, and are thought to represent the most massive early type galaxies, many of which are classified as `Brightest Cluster Galaxies'. Fig.~\ref{Hubble} shows that host galaxies of GPS radio sources have similar optical luminosities to LRGs, indicating that powerful young radio sources are hosted by the most massive early type galaxies. Note that the flattening of the distribution of LRGs at low redshift is probably due to the known problem that SDSS subtracts too high background levels for large sources, which can result in a magnitude difference of up to $\sim$1 magnitude.
\subsection{Absolute magnitudes} \label{SSabsmag}
To study the cosmological evolution of GPS galaxies in more detail, we determined their absolute magnitudes $M_{R_C}$ as a function of redshift. These can be calculated from the apparent magnitudes $m_{R_C}$, using (Hogg \cite{Hogg}):
\begin{eqnarray} \label{EqMr}
  m_{R_C} & = & M_{R_C} + \mu + K \nonumber \\
          & = & M_{R_C} + 5\log\left(\frac{D_L}{10\ pc}\right) - 2.5
       \log\left[\frac{L_{\lambda/(1+z)}}{(1+z)\ L_{\lambda}}\right]
\end{eqnarray}
with $\mu$ the distance modulus, which depends, through the luminosity distance $D_L$ , on the redshift of the source and on the assumed cosmology (we adopted the cosmological parameters as found by WMAP; Spergel et al. \cite{Spergel}). The last term, $K$, is the `k-correction', which is required to convert the absolute magnitude to rest-frame $R$-band. This depends on the assumed spectral energy distribution (SED) of the source and is represented by the luminosity at the intrinsic wavelength $L_{\lambda/(1+z)}$ divided by that at the observed wavelength $L_{\lambda}$.

All variables in equation (\ref{EqMr}) are determined in a straightforward way from the observations and adopted cosmology, except for the k-correction. We therefore determined $m_R-\mu$ for each galaxy in the sample and determined a range of possible k-corrections, depending on the age and metallicity of the stellar population. For these we used the model galaxy SEDs by Worthey (\cite{Worthey}). In figure (\ref{FigEvol}) we show $m_R-\mu$ of the combined samples (see Sect.~\ref{SSHub}) of GPS galaxies (solid squares) and 3C galaxies (open circles) as a function of redshift. The three solid lines show the expected trend in $m_R-\mu$ for a stellar population with a formation redshift $z_f$ = 1.5 and metallicities [Fe/H] of 0.3, 0.0, and -0.225 compared to solar. In figure (3) we show the same data, overplotted with the k-correction for a stellar population with solar metallicity and a range of formation redshifts of $z_f$ = 1.5, 2.0, and 3.0.

These figures show that the data are in best agreement with a recent formation redshift of $1.5 - 2.0$ (with a possible range of metallicities). However, we do not believe that the host galaxies really are that young, since GPS galaxies are always classified as early type galaxies, which are generally thought to have formed at higher redshifts ($z \sim 5$). In addition, Fig. \ref{Hubble} also indicates that GPS galaxies are old, massive, early type galaxies. Therefore we interpret this result as evidence for starburst activity and/or AGN induced light. We note that, although the optical/UV contribution from the alignment effect appears significantly smaller in GPS galaxies than for classical 3C radio galaxies, it does not exclude such AGN induced light being present at a low level. This extra blue light would brighten galaxies the most at high redshifts, where the $R$-band probes the rest-frame UV, and therefore could mimic a young stellar population and low formation redshifts. A thorough investigation using deep optical and infrared spectra will be needed to determine the possible contributions from the AGN and young starbursts to the overall galaxy spectrum.
\begin{acknowledgements}
This research has made use of observations collected at the ESO/Paranal Very Large Telescope and the ESO/La Silla 3.6m Telescope, and made use of the VizieR catalogue access tool, CDS, Strasbourg, France (Ochsenbein et al. \cite{Ochsenbein}).
This publication makes use of data products from the USNO-B1.0 Catalog and the Two Micron All Sky Survey, which is a joint project of the University of Massachusetts and the Infrared Processing and Analysis Center/California Institute of Technology, funded by the National Aeronautics and Space Administration and the National Science Foundation.
\end{acknowledgements}
{}
\appendix
\section{Comments on selected sources}
\textbf{PKS~J0108$-$1201} The carbon line CIII] 1909 \AA\ and the neon lines ([NeV] 3346 \AA, [NeV] 3426 \AA) are detected. By smoothing the spectrum with a boxcar of 20 pixels the MgII 2799 \AA\ line is revealed as a broad (FWHM $\sim$ 250 \AA\ $\sim$ 10$^4$ km/s) spectral feature, centered at $\sim$ 7100 \AA.\\
\textbf{PKS~J0210+0419} Neither a continuum nor emission lines were detected for this object. This is consistent with the non-detection in the $R$-band (Snellen et al. \cite{Snellen02}).\\
\textbf{PKS~J0401$-$2921} The spectrum shows very strong oxygen lines, weaker neon lines and the CaII K and H absorption features with a 4000 \AA\ break.\\
\begin{figure}[b]
\centering
 \includegraphics[width=8.5cm]{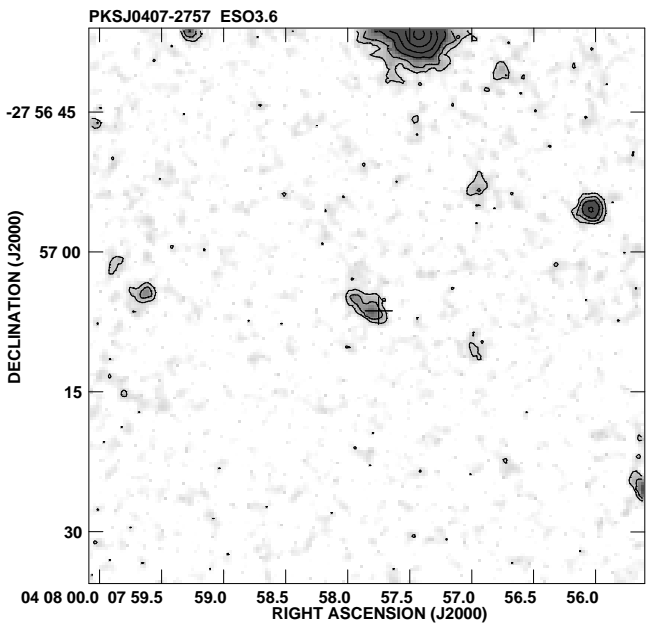}
\caption{$R$-band image of PKS~J0407$-$2757 taken from Snellen et al. (\cite{Snellen02}). The GPS radio source is located in the South Eastern object, as indicated with the plus sign. Most of the line emission originates from the object North Western of the radio source.}
\label{PKSJ0407-2757}
\end{figure}
\textbf{PKS~J0407$-$2757} The slit was tilted 45\degr\ so that the object North Western of the radio source (separation $\sim$ 2.5\arcsec) was included. The North Western object shows many strong emission lines, where the South Eastern object shows [OII] 3727 \AA\ and very weak neon lines. An $R$-band image of this source is shown in Fig. \ref{PKSJ0407-2757}.\\
\textbf{PKS~J0433$-$0229} The spectrum shows [OII] 3727 \AA\ in emission and the Ca II H, G-band, H$_{\gamma}$, H$_{\beta}$ and MgI features in absorption.\\
\textbf{PKS~J1057+0012} This object was observed in photometric (1200 s) and spectroscopic (2700 s) mode on the ESO 3.6m Telescope. The images resulted in a new identification with a $m_{R_C}$ = 22.5 $\pm$ 0.3 object, but the spectrum was not deep enough to determine a redshift for this source.\\
\textbf{PKS~J1109+1043} This radio source has erroneously been identified with a $m_{R_C}$ = 20.5 object. It was already noted (Snellen et al. 2002) that there was a large offset between the radio and optical position. Our observations resulted in an identification with a much fainter object ($m_{R_C}$ = 22.6 $\pm$ 0.3) at the radio position. The S/N of the spectrum (2700 s on the 3.6m Telescope) was too low to see spectral features.\\
\textbf{PKS~J1135$-$0021} This radio source has erroneously been identified with a star ($m_{R_C}$ = 16.5). In our long slit spectrum we found, at a distance of about 2.4\arcsec from the typical stellar spectrum, a faint AGN spectrum from which we determined a redshift of 0.975. From the acquisition image we also estimated the magnitude of the object: $m_{R_C}$ = 21.9 $\pm$ 0.4. \\
\textbf{PKS~J1203+0414} Spectroscopic observations of this object show a typical quasar spectrum with broad carbon and magnesium lines and the object should therefore be removed from the sample of GPS galaxies.\\
\textbf{PKS~J1345$-$3015} The spectrum shows a strong [OII] 3727 \AA\ line, the CII] 2326 \AA, [NeV] 3426 \AA\ and [NeIII] 3869 \AA\ lines and a very broad MgII feature.\\
\textbf{PKS~J1352+1107} The [OII] 3727 \AA\ line, the neon emission lines and the CaII K and H absorption features are detected.\\
\textbf{PKS~J1506$-$0919} Spectroscopic observations of this object show a typical quasar spectrum with broad carbon and magnesium lines and the object should therefore be removed from the sample of GPS galaxies.\\
\textbf{PKS~J1556$-$0622} Only one emission line is detected, so only a tentative redshift could be determined. We assumed the line to be [OII] 3727 \AA, because of the resemblance between this spectrum and those of PKS~J2339$-$0604 and PKS~J2212+0152, and the absence of viable alternatives.\\
\textbf{PKS~J1648+0242} The spectrum shows very weak [OII] 3727 \AA\ and neon emission lines, but strong CaII K and H absorption features, the G-band and the 4000 \AA\ break.\\
\textbf{PKS~J1734+0926} The [OII] 3727 \AA\ and [NeIII] 3869 \AA\ emission lines are weak, but the H$_{\gamma}$ and CaII K and H absorption features, the G-band and the 4000 \AA\ break are strong.\\
\textbf{PKS~J2212+0152} The [OII] 3727 \AA\ emission line is strong and the MgII 2798 \AA\ and neon lines are weak.\\
\textbf{PKS~J2339$-$0604} The spectrum shows a strong [OII] 3727 \AA\ line and weak CII] 2326 \AA, MgII 2798 \AA\ and [NeIII] 3869 \AA\ lines.\\

\begin{figure*}
\centering
{\sl \section{Tables and figures }}
 \includegraphics[width=17cm]{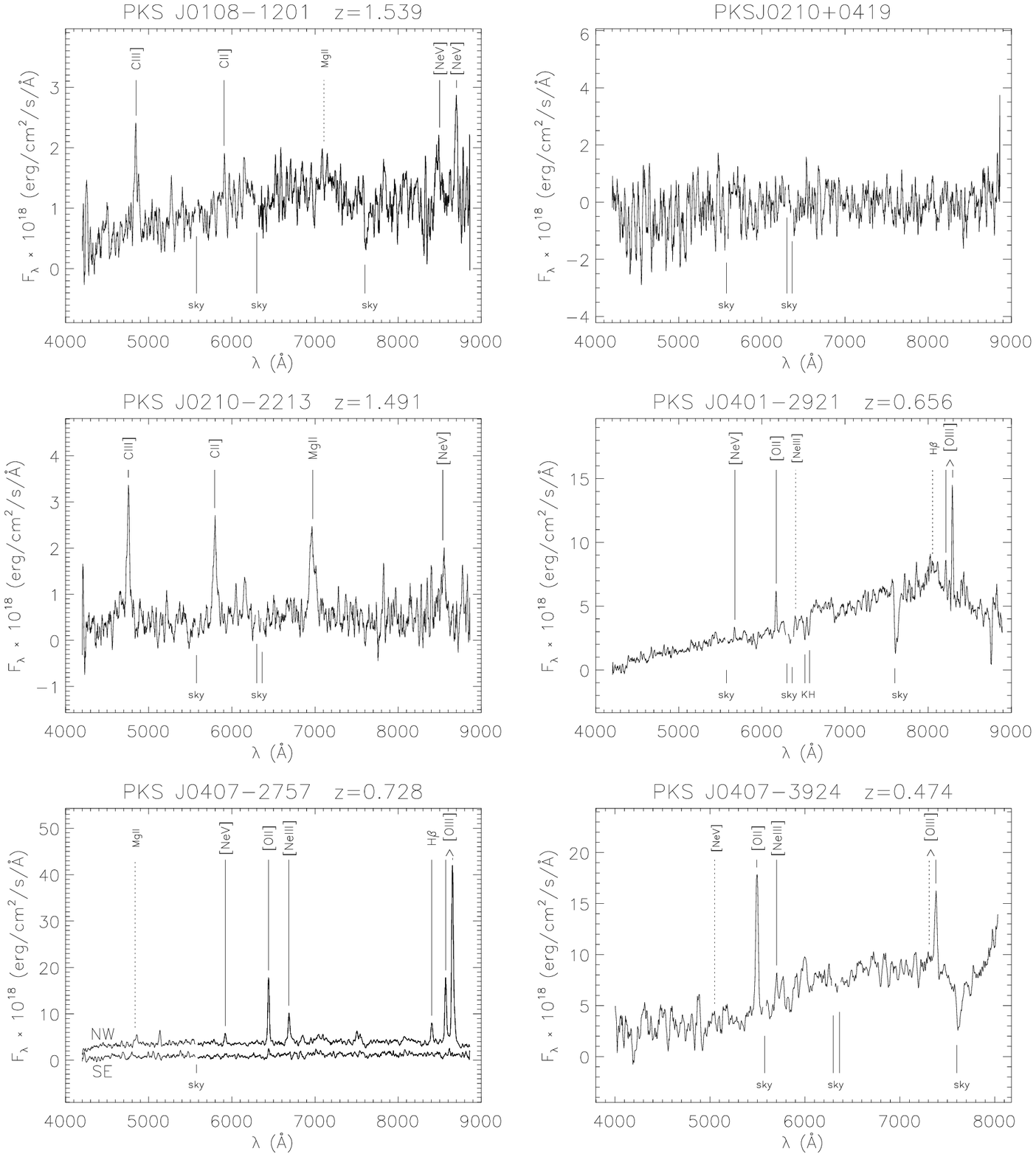}
 \caption{Spectra from the ESO Very Large Telescope and ESO 3.6m Telescope of 24 objects in the sample. From 21 spectra, a redshift could be deduced, which is indicated at the top. Labeled are emission and absorption lines that were used to determine the redshift (drawn lines) and wavelengths where other common emission lines are expected (dotted lines). Regions where the night sky emission lines are strong have been omitted. These regions are indicated, as well as the 7600 \AA\ sky absorption feature.}
 \label{Figspec}
\end{figure*}
\addtocounter{figure}{-1}
\begin{figure*}
\centering
 \includegraphics[width=17cm]{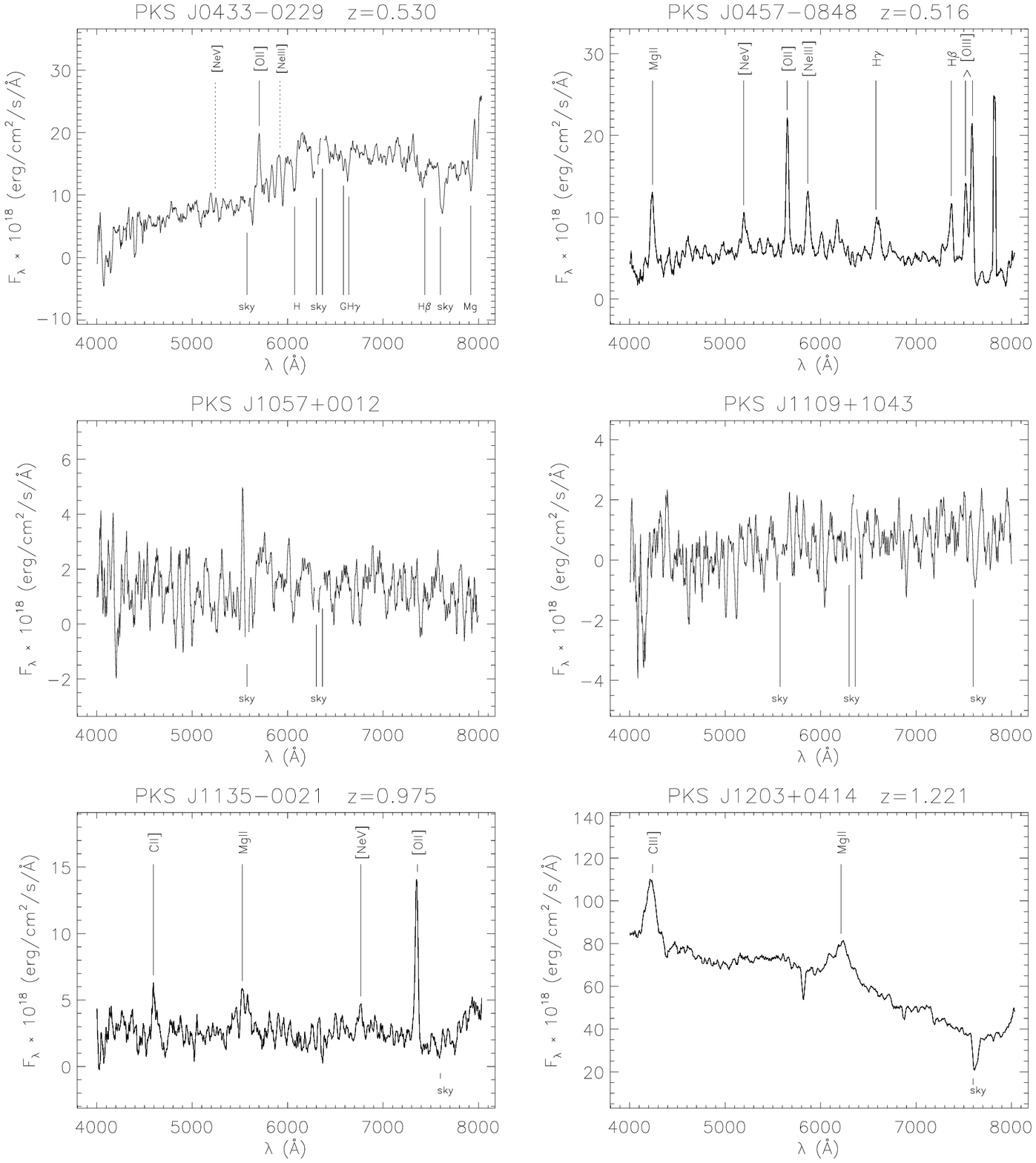}
 \caption{{\it continued}}
\end{figure*}
\addtocounter{figure}{-1}
\begin{figure*}
\centering
 \includegraphics[width=17cm]{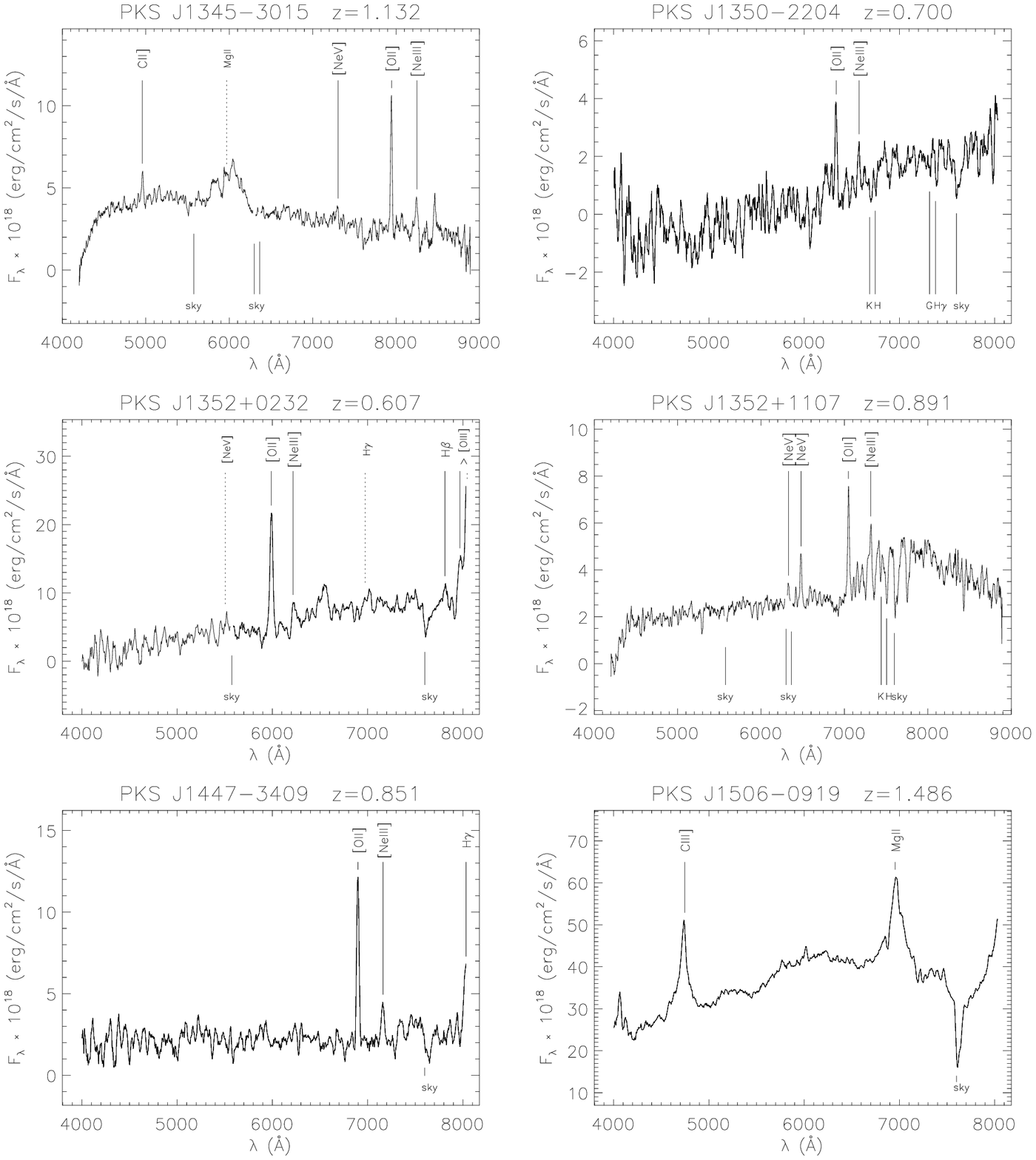}
 \caption{{\it continued}}
\end{figure*}
\addtocounter{figure}{-1}
\begin{figure*}
\centering
 \includegraphics[width=17cm]{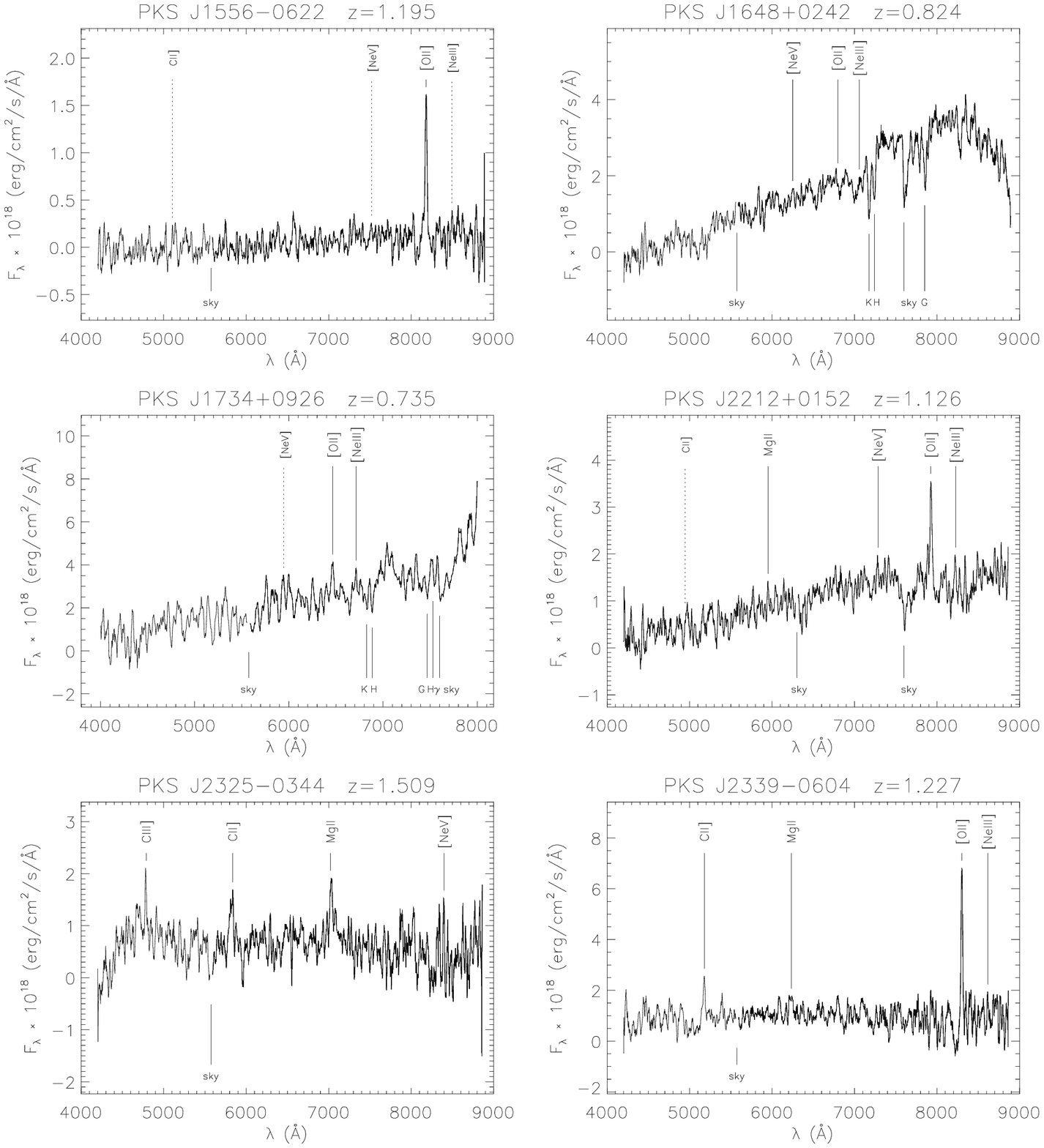}
 \caption{{\it continued}}
\end{figure*}
\begin{table*}
\caption{The radio and optical properties of objects in the southern/equatorial sample of GPS galaxies.}
\label{sample}
\begin{tabular}{c c c c c c c c c c c c}
\hline \hline
IAU		& Other		& \multicolumn{2}{c}{Radio position}	& $m_{R_C}$		& z	& $S_{\rm 2.7GHz}$	& $\nu_{\rm peak}$	& $S_{\rm peak}$	& Ref.	& Ref.	& Comm. \\
name		& name		& \multicolumn{2}{c}{(J2000)}		& (mag)			&	& (Jy)	& (GHz)	& (Jy)	& $m_R$	& z	& \\
\hline
J0022+0014	& 4C+00.02	& 00 22 25.48	& + 00 14 56.0	& 18.10 $\pm$ 0.20	& 0.305	& 1.94	& 0.6	& 3.1	& 3	& 3	& A \\
J0108$-$1201	& B0105$-$122	& 01 08 13.20	& $-$12 00 50.3	& 22.39 $\pm$ 0.06	& 1.539	& 0.52	& 1.0	& 0.9	& 2	& 1	& \\
J0206$-$3024	& B0204$-$306	& 02 06 43.26	& $-$30 24 58.2	& 21.00 $\pm$ 0.50	& 	& 0.58	& 0.5	& 0.9	& 14	&	& \\
J0210+0419	& B0208+040	& 02 10 44.52	& +04 19 35.4	& $K_s$ 18.3 $\pm$ 0.2		& 	& 0.56	& 0.4	& 1.3	& 1	&	& \\
J0210$-$2213	& B0207$-$224	& 02 10 10.05	& $-$22 13 36.6	& 23.52 $\pm$ 0.13	& 1.491	& 0.86	& 1.5	& 1.1	& 2	& 1	& \\
J0242$-$2132	& B0240$-$217	& 02 42 35.87	& $-$21 32 26.2	& 17.10 $\pm$ 0.50	& 0.314	& 0.97	& 1.0	& 1.3	& 14	& 12	& \\
J0323+0534	& 4C+05.14	& 03 23 20.27	& +05 34 11.9	& 19.20 $\pm$ 0.50	& 	& 1.60	& 0.4	& 7.1	& 14	&	& \\
J0401$-$2921	& B0359$-$294	& 04 01 21.50	& $-$29 21 26.1	& 21.00 $\pm$ 0.20	& 0.656	& 0.58	& 0.4	& 1.0	& 1	& 1	& \\
J0407$-$3924	& B0405$-$395	& 04 07 34.43	& $-$39 24 47.2	& 20.40 $\pm$ 0.50	& 0.474	& 0.52	& 0.4	& 1.4	& 14	& 1	& \\
J0407$-$2757	& B0405$-$280	& 04 07 57.94	& $-$27 57 05.1	& 21.14 $\pm$ 0.04	& 0.728	& 0.93	& 1.5	& 1.4	& 2	& 1	& \\
J0433$-$0229	& 4C$-$02.17	& 04 33 54.90	& $-$02 29 56.0	& 19.10 $\pm$ 0.50	& 0.530	& 1.04	& 0.4	& 3.0	& 14	& 1	& \\
J0441$-$3340	& B0439$-$337	& 04 41 33.80	& $-$33 40 03.6	& 22.50 $\pm$ 0.40	& 	& 0.88	& 1.5	& 1.2	& 1	&	& \\
J0457$-$0848	& B0454$-$088	& 04 57 20.24	& $-$08 49 05.2	& 20.30 $\pm$ 0.50	& 0.516	& 0.58	& 0.4	& 1.0	& 14	& 1	& \\
J0503+0203	& B0500+019	& 05 03 21.20	& +02 03 04.6	& i21.0 $\pm$ 0.20		& 0.583	& 2.46	& 2.5	& 2.5	& 5	& 5	& A \\
J0943$-$0819	& B0941$-$080	& 09 43 36.86	& $-$08 19 32.0	& 17.50 $\pm$ 0.20	& 0.228	& 1.73	& 0.4	& 4.2	& 10	& 11	& A \\
J0913+1454	& B0910+151	& 09 13 35.00	& +14 54 20.1	& 22.90 $\pm$ 0.50	& 	& 0.54	& 0.6	& 1.1	& 1	&	& \\
J1044$-$2712	& B1042$-$269	& 10 44 37.63	& $-$27 12 18.6	& 22.80 $\pm$ 0.40	& 	& 0.55	& 1.5	& 0.8	& 1	&	& \\
J1057+0012	& B1054+004	& 10 57 15.78	& +00 12 03.7	& 22.30 $\pm$ 0.30	&	& 0.58	& 0.4	& 1.6	& 1	&	& \\
J1109+1043	& B1107+109	& 11 09 46.04	& +10 43 43.4	& 22.60 $\pm$ 0.30	& 	& 0.80	& 0.5	& 2.4	& 1	&	& \\
J1110$-$1858	& B1107$-$187	& 11 10 00.45	& $-$18 58 49.2	& 19.60 $\pm$ 0.20	& 0.497	& 0.65	& 1.0	& 0.9	& 9	& 13	& \\
J1120+1420	& 4C+14.41	& 11 20 27.81	& +14 20 55.0	& 20.10 $\pm$ 0.10	& 0.362	& 1.50	& 0.4	& 3.7	& 5	& 5	& A \\
J1122$-$2742	& B1120$-$274	& 11 22 56.41	& $-$27 42 48.2	& 23.00 $\pm$ 0.50	& 	& 0.74	& 1.4	& 0.8	& 1	&	& \\
J1135$-$0021	& 4C$-$00.45	& 11 35 12.96	& $-$00 21 19.5	& 21.90 $\pm$ 0.40	& 0.975	& 0.76	& 0.4	& 2.9	& 1	& 1	& \\
J1203+0414	& B1200+045	& 12 03 21.95	& +04 14 17.7	& 18.80 $\pm$ 0.50	& 1.221	& 0.52	& 0.4	& 1.4	& 14	& 1	& E \\
J1345$-$3015	& B1343$-$300	& 13 45 51.52	& $-$30 15 04.1	& 21.20 $\pm$ 0.20 & 1.132	& 0.56	& 0.4	& 2.5	& 1	& 1	& \\
J1347+1217	& 4C+12.50	& 13 47 33.36	& +12 17 24.2	& 15.20 $\pm$ 0.20	& 0.122	& 3.88	& 0.4	& 8.8	& 3	& 3	& A \\
J1350$-$2204	& B1347$-$218	& 13 50 14.33	& $-$22 04 43.7	& 20.93 $\pm$ 0.05	& 0.700	& 0.72	& 0.4	& 1.4	& 2	& 1	& \\
J1352+0232	& B1349+027	& 13 52 30.68	& +02 32 47.7	& 20.00 $\pm$ 0.50	& 0.607	& 0.78	& 0.4	& 2.0	& 14	& 1	& \\
J1352+1107	& 4C+11.46	& 13 52 56.37	& +11 07 07.7	& 21.00 $\pm$ 0.20 & 0.891	& 0.78	& 0.4	& 3.6	& 1	& 1	& \\
J1447$-$3409	& B1444$-$339	& 14 47 19.69	& $-$34 09 16.2	& 21.00 $\pm$ 0.10	& 0.851	& 0.50	& 0.5	& 1.0	& 14	& 1	& \\
J1506$-$0919	& B1503$-$091	& 15 06 03.05	& $-$09 19 12.5	& 19.70 $\pm$ 0.50	& 1.486	& 0.87	& 0.6	& 1.6	& 14	& 1	& E \\
J1521+0430	& 4C+04.51	& 15 21 14.51	& +04 30 20.0	& 22.10 $\pm$ 0.11	& 1.296	& 2.30	& 1.0	& 5.4	& 7	& 11	& A \\
J1543$-$0757	& B1540$-$077	& 15 43 01.69	& $-$07 57 03.6	& 17.40 $\pm$ 0.10	& 0.172	& 1.21	& 0.4	& 1.7	& 6	& 6	& C \\
J1546+0026	& B1543+005	& 15 46 09.50	& +00 26 24.6	& 20.10 $\pm$ 0.20	& 0.556	& 1.24	& 0.6	& 2.2	& 3	& 6	& C \\
J1548$-$1213	& B1545$-$120	& 15 48 12.97	& $-$12 13 31.8	& 21.88 $\pm$ 0.13	& 0.883	& 1.45	& 0.4	& 3.7	& 2	& 2	& \\
J1556$-$0622	& 4C$-$06.43	& 15 56 13.99	& $-$06 22 37.8	& 22.20 $\pm$ 0.13	& (1.195)	& 0.77	& 0.4	& 2.4	& 2	& 1	& \\
J1600$-$0037	& B1557$-$004	& 16 00 00.91	& $-$00 37 23.3	& $-$			& 	& 0.54	& 1.0	& 1.2	& 2	&	& D \\
J1604$-$2223	& B1601$-$222	& 16 04 01.45	& $-$22 23 41.3	& 18.75 $\pm$ 0.10	& 0.141	& 0.57	& 0.6	& 1.0	& 6	& 2	& \\
J1640+1220	& 4C+12.60	& 16 40 47.96	& +12 20 02.1	& 21.36 $\pm$ 0.20	& 1.150	& 1.48	& 0.4	& 3.7	& 2	& 2	& \\
J1648+0242	& 4C+02.43	& 16 48 31.79	& +02 42 46.0	& 22.10 $\pm$ 0.30		& 0.824	& 0.61	& 0.4	& 3.4	& 1	& 1	& \\
J1734+0926	& B1732+094	& 17 34 58.38	& +09 26 57.8	& 20.80 $\pm$ 0.10	& 0.735	& 1.08	& 1.0	& 1.1	& 6	& 1	& B \\
J2011$-$0644	& B2008$-$068	& 20 11 14.22	& $-$06 44 03.6	& 21.18 $\pm$ 0.04	& 0.547	& 2.09	& 1.4	& 2.6	& 2	& 2	& A \\
J2058+0540	& 4C+05.78	& 20 58 28.84	& +05 42 50.7	& 23.40 $\pm$ 0.30	& 1.381	& 0.65	& 0.4	& 3.1	& 8	& 8	& \\
J2123$-$0112	& B2121$-$014	& 21 23 39.12	& $-$01 12 34.3	& 23.30 $\pm$ 0.10	& 1.158	& 0.64	& 0.4	& 2.0	& 4	& 3	& \\
J2130+0502	& B2128+048	& 21 30 32.88	& +05 02 17.5	& 22.21 $\pm$ 0.07	& 0.990	& 3.12	& 1.0	& 4.8	& 2	& 11	& A \\
J2151+0552	& B2149+056	& 21 51 37.88	& +05 52 13.0	& 20.20 $\pm$ 0.20	& 0.740	& 1.01	& 5.0	& 1.2	& 4	& 11	& A, E \\
J2212+0152	& 4C+01.69	& 22 12 37.97	& +01 52 51.7	& i22.0 $\pm$ 0.20		& 1.126	& 1.80	& 0.4	& 4.6	& 5	& 1	& A \\
J2325$-$0344	& B2322$-$040	& 23 25 10.23	& $-$03 44 46.7	& 23.50 $\pm$ 0.20	& 1.509	& 0.91	& 1.4	& 1.2	& 10	& 1	& B \\
J2339$-$0604	& 4C$-$06.76	& 23 37 11.95	& $-$06 04 12.4	& 22.91 $\pm$ 0.20	& 1.227	& 0.80	& 0.4	& 3.8	& 2	& 1	& \\
\hline
\end{tabular}
 \\
 \\
{\sl Comments:} (A) also in the sample of $>$ 1 Jy GPS sources from Stanghellini et al. (\cite{Stanghellini98}); (B) also in the sample of de Vries et al. (\cite{Vries97}) (C) also in the sample of de Vries et al. (\cite{Vries00}); (D) near a bright star, no magnitude; (E) radio source is optically identified with a quasar; will be excluded from the sample.\\
{\sl References:} (1) this paper; (2) Snellen et al. (\cite{Snellen02}); (3) Snellen et al. (\cite{Snellen96a}); (4) O'Dea, Baum \& Morris (\cite{O'Dea90}); (5) de Vries et al. (\cite{Vries95}); (6) de Vries et al. (\cite{Vries00}). (7) Biretta, Schneider \& Gunn (\cite{Biretta}); (8) Stern et al. (\cite{Stern}); (9) Fugmann, Meisenheimer \& Roeser (\cite{Fugmann}); (10) Stanghellini et al. (\cite{Stanghellini93}); (11) Stanghellini et al. (\cite{Stanghellini98}); (12) Otrupcek \& Wright (\cite{Otrupcek}); (13) Drinkwater et al. (\cite{Drinkwater}); (14) Digitized Sky Survey II; APM catalogue (Irwin et al. \cite{Irwin}); SuperCosmos Sky Surveys (Hambly et al. \cite{Hambly}).
\end{table*}
\begin{longtable}{c c c c c c c}
\caption{\label{specdet} Details of the spectroscopic observations.}\\
\hline\hline
 & & & \multicolumn{3}{c}{Spectral feature} & \\
Object name & t & Telescope & Species & $\lambda_{rest}$ & $\lambda_{obs}$ & z \\
 & (s) & & & (\AA) & (\AA) & \\
\hline
\endfirsthead
\caption{Details of the spectroscopic observations, continued.}\\
\hline\hline
 & & & \multicolumn{3}{c}{Spectral feature} & \\
Object name & t & Telescope & Species & $\lambda_{rest}$ & $\lambda_{obs}$ & z \\
 & (s) & & & (\AA) & (\AA) & \\
\hline
\endhead
\hline
\endfoot
   PKS~J0108$-$1201 & 1800 & ESO VLT & & & & 1.539 $\pm$ 0.001 \\
          & & & CIII]    & 1909 & 4845 & 1.538 \\
          & & & CII]     & 2326 & 5911 & 1.541 \\
          & & & [NeV]    & 3346 & 8488 & 1.537 \\
          & & & [NeV]    & 3426 & 8700 & 1.539 \\
   PKS~J0210+0419 & 1800 & ESO VLT & & & & \\
   PKS~J0210$-$2213 & 1800 & ESO VLT & & & & 1.491 $\pm$ 0.003 \\
          & & & CIII]    & 1909 & 4757 & 1.492 \\
          & & & CII]     & 2326 & 5799 & 1.493 \\
          & & & MgII     & 2798 & 6964 & 1.489 \\
          & & & [NeV]    & 3426 & 8553 & 1.497 \\
   PKS~J0401$-$2921 & 1800 & ESO VLT & & & & 0.656 $\pm$ 0.001 \\
          & & & [NeV]    & 3426 & 5672 & 0.656 \\
          & & & [OII]    & 3727 & 6170 & 0.656 \\
          & & & CaII K   & 3934 & 6518 & 0.657 \\
          & & & CaII H   & 3969 & 6562 & 0.653 \\
          & & & [OIII]   & 4959 & 8212 & 0.656 \\
          & & & [OIII]   & 5007 & 8290 & 0.656 \\
   PKS~J0407$-$2756 & 1800 & ESO VLT & & & & 0.728 $\pm$ 0.001 \\
   North West & & & [NeV]    & 3426 & 5920 & 0.728 \\
          & & & [OII]    & 3727 & 6442 & 0.728 \\
          & & & [NeIII]  & 3869 & 6687 & 0.728 \\
          & & & H$\beta$ & 4861 & 8404 & 0.729 \\
          & & & [OIII]   & 4959 & 8571 & 0.728 \\
          & & & [OIII]   & 5007 & 8654 & 0.728 \\
   PKS~J0407$-$2756 & 1800 & ESO VLT & & & & 0.729 $\pm$ 0.001 \\
   South East & & & [NeV]    & 3426 & 5929 & 0.731 \\
          & & & [OII]    & 3727 & 6442 & 0.729 \\
   PKS~J0407$-$3924 & 2700 & ESO 3.6m & & & & 0.474 $\pm$ 0.001 \\
          & & & [OII]    & 3727 & 5496 & 0.475 \\
          & & & [NeIII]  & 3869 & 5700 & 0.473 \\
          & & & [OIII]   & 5007 & 7379 & 0.474 \\
   PKS~J0433$-$0229 & 1200 & ESO 3.6m & & & & 0.530 $\pm$ 0.001 \\
          & & & [OII]       & 3727 & 5496 & 0.530 \\
          & & & CaII H$_a$  & 3969 & 6071 & 0.530 \\
          & & & G-band$_a$  & 4304 & 6584 & 0.530 \\
          & & & H$\gamma_a$ & 4340 & 6634 & 0.529 \\
          & & & H$\beta_a$  & 4861 & 7411 & 0.525 \\
          & & & MgI$_a$     & 5175 & 7921 & 0.531 \\
   PKS~J0457$-$0848 & 2700 & ESO 3.6m & & & & 0.516 $\pm$ 0.001 \\
          & & & MgII      & 2798 & 4238 & 0.515 \\
          & & & [NeV]     & 3426 & 5195 & 0.516 \\
          & & & [OII]     & 3727 & 5653 & 0.517 \\
          & & & [NeIII]   & 3869 & 5865 & 0.516 \\
          & & & [SII]     & 4072 & 6174 & 0.516 \\
          & & & H$\gamma$ & 4340 & 6582 & 0.517 \\
          & & & [OIII]    & 4363 & 6618 & 0.517 \\
          & & & H$\beta$  & 4861 & 7373 & 0.517 \\
          & & & [OIII]    & 4959 & 7521 & 0.517 \\
          & & & [OIII]    & 5007 & 7589 & 0.516 \\
   PKS~J1057+0012 & 2700 & ESO 3.6m & & & & \\
   PKS~J1109+1043 & 2700 & ESO 3.6m & & & & \\
   PKS~J1135$-$0021 & 1800 & ESO 3.6m & & & & 0.975 $\pm$ 0.002 \\
          & & & CII]      & 2326 & 4593 & 0.975 \\
          & & & MgII      & 2798 & 5531 & 0.977 \\
          & & & [NeV]     & 3426 & 6769 & 0.976 \\
          & & & [OII]     & 3727 & 7351 & 0.972 \\
   PKS~J1203+0414 & 1200 & ESO 3.6m & & & & 1.221 $\pm$ 0.004 \\
          & & & CIII]    & 1909 & 4230 & 1.216 \\
          & & & MgII     & 2798 & 6225 & 1.225 \\
   PKS~J1345$-$3015 & 1800 & ESO VLT & & & & 1.132 $\pm$ 0.001 \\
          & & & CII]     & 2326 & 4960 & 1.132 \\
          & & & [NeV]    & 3426 & 7297 & 1.130 \\
          & & & [OII]    & 3727 & 7944 & 1.132 \\
          & & & [NeIII]  & 3869 & 8248 & 1.132 \\
   PKS~J1350$-$2204 & 2700 & ESO 3.6m & & & & 0.700 $\pm$ 0.002 \\
          & & & [OII]       & 3727 & 6333 & 0.699 \\
          & & & [NeIII]     & 3869 & 6572 & 0.699 \\
          & & & CaII K$_a$  & 3934 & 6707 & 0.705 \\
          & & & CaII H$_a$  & 3969 & 6752 & 0.701 \\
          & & & G-band$_a$  & 4304 & 7317 & 0.700 \\
          & & & H$\gamma_a$ & 4340 & 7390 & 0.703 \\
   PKS~J1352+0232 & 1800 & ESO 3.6m & & & & 0.607 $\pm$ 0.001 \\
          & & & [OII]     & 3727 & 5990 & 0.607 \\
          & & & [NeIII]   & 3869 & 6215 & 0.606 \\
          & & & H$\beta$  & 4861 & 7813 & 0.607 \\
          & & & [OIII]    & 4959 & 7968 & 0.607 \\
   PKS~J1352+1107 & 1800 & ESO VLT & & & & 0.891 $\pm$ 0.001 \\
          & & & [NeV]    & 3346 & 6329 & 0.892 \\
          & & & [NeV]    & 3426 & 6478 & 0.891 \\
          & & & [OII]    & 3727 & 7050 & 0.892 \\
          & & & [NeIII]  & 3869 & 7319 & 0.892 \\
          & & & CaII K   & 3934 & 7434 & 0.890 \\
          & & & CaII H   & 3969 & 7508 & 0.892 \\
   PKS~J1447$-$3409 & 5400 & ESO 3.6m & & & & 0.851 $\pm$ 0.001 \\
          & & & [OII]     & 3727 & 6898 & 0.851 \\
          & & & [NeIII]   & 3869 & 7161 & 0.851 \\
          & & & H$\gamma$ & 4340 & 8047 & 0.854 \\
   PKS~J1506$-$0919 & 2700 & ESO 3.6m & & & & 1.486 $\pm$ 0.003 \\
          & & & CIII]    & 1909 & 4740 & 1.483 \\
          & & & MgII     & 2798 & 6966 & 1.490 \\
   PKS~J1556$-$0622 & 1800 & ESO VLT & & & & 1.195 $\pm$ 0.001 \\
          & & & [OII]    & 3727 & 8182 & 1.195 \\
   PKS~J1648+0242 & 1800 & ESO VLT & & & & 0.824 $\pm$ 0.001 \\
          & & & [NeV]    & 3426 & 6252 & 0.825 \\
          & & & [OII]    & 3727 & 6777 & 0.818 \\
          & & & [NeIII]  & 3869 & 7069 & 0.827 \\
          & & & CaII K   & 3934 & 7176 & 0.824 \\
          & & & CaII H   & 3969 & 7242 & 0.825 \\
          & & & G-band   & 4305 & 7856 & 0.825 \\
   PKS~J1734+0926 & 5400 & ESO 3.6m & & & & 0.735 $\pm$ 0.001 \\
          & & & [NeV]      & 3426 & 5938 & 0.733 \\
          & & & [OII]      & 3727 & 6468 & 0.735 \\
          & & & [NeIII]    & 3869 & 6713 & 0.735 \\
          & & & CaII K$_a$ & 3934 & 6829 & 0.736 \\
          & & & CaII H$_a$ & 3969 & 6884 & 0.735 \\
   PKS~J2212+0152 & 1800 & ESO VLT & & & & 1.126 $\pm$ 0.001 \\
          & & & MgII     & 2798 & 5955 & 1.128 \\
          & & & [NeV]    & 3426 & 7280 & 1.125 \\
          & & & [OII]    & 3727 & 7926 & 1.127 \\
          & & & [NeIII]  & 3869 & 8219 & 1.124 \\
   PKS~J2325$-$0344 & 1800 & ESO VLT & & & & 1.509 $\pm$ 0.002 \\
          & & & CIII]    & 1909 & 4781 & 1.505 \\
          & & & CII]     & 2326 & 5837 & 1.510 \\
          & & & MgII     & 2798 & 7029 & 1.512 \\
          & & & [NeV]    & 3346 & 8394 & 1.109 \\
   PKS~J2339$-$0604 & 3600 & ESO VLT & & & & 1.227 $\pm$ 0.001 \\
          & & & CII]     & 2326 & 5181 & 1.226 \\
          & & & MgII     & 2798 & 6230 & 1.227 \\
          & & & [OII]    & 3727 & 8300 & 1.227 \\
          & & & [NeIII]  & 3869 & 8617 & 1.226 \\
\end{longtable}
\end{document}